\documentclass[a4paper,aps,superscriptaddress,floatfix,nofootinbib,twocolumn,longbibliography]{revtex4-1}

\usepackage{amsmath, amsthm, amssymb} 
\usepackage{algpseudocode} 
\usepackage{comment} 
\usepackage{graphicx}
\usepackage{dcolumn}
\usepackage{bm}
\usepackage{hyperref}
\usepackage[mathlines]{lineno}

\usepackage{multirow}

\theoremstyle{definition}

\usepackage{color}

\begin{document}
\title{
Multiplex reconstruction with partial information
}

\author{Daniel Kaiser}
\thanks{These authors contributed equally to this work.}
   \affiliation{Center for Complex Networks and Systems Research, Luddy School
  of Informatics, Computing, and Engineering, Indiana University, Bloomington,
  Indiana 47408, USA}

\author{Siddharth Patwardhan}
\thanks{These authors contributed equally to this work.}
    \affiliation{Center for Complex Networks and Systems Research, Luddy School
  of Informatics, Computing, and Engineering, Indiana University, Bloomington,
  Indiana 47408, USA}

\author{Filippo Radicchi}
\affiliation{Center for Complex Networks and Systems Research, Luddy School
  of Informatics, Computing, and Engineering, Indiana University, Bloomington,
  Indiana 47408, USA}
\email{filiradi@indiana.edu}

\begin{abstract}
A multiplex is a collection of network layers, each representing a specific type of edges. This appears to be a genuine representation for many real-world systems.
	However,
	due to a variety of potential factors, such as limited budget and equipment, or physical impossibility, multiplex data can be difficult to observe directly. Often, only partial information on the layer structure of 
	the system is available, whereas the remaining information is
	in the form of a single-layer network. 
	In this work, we face the problem of 
	reconstructing the hidden multiplex structure of an aggregated network from partial information.
	We propose an algorithm that leverages the layer-wise community structure that can be learned from partial observations to reconstruct the ground-truth
	 topology of the unobserved part of the multiplex.
	The algorithm is characterized by a computational time that grows linearly with the network size. We perform  a systematic study of reconstruction problems for both synthetic and real-world multiplex networks.
	We show that the ability of the proposed method to solve the reconstruction problem is affected by the heterogeneity of the individual layers and the similarity among the layers.
	On real-world networks, we observe that the accuracy of the reconstruction saturates quickly as the amount of available information increases. In genetic interaction and scientific collaboration multiplexes for example, we find that $10\%$ of ground-truth information yields $70\%$ accuracy, while $30\%$ information allows for more than $90\%$ accuracy. 
\end{abstract}

\maketitle


\section{Introduction}
 Networks have emerged as powerful modeling frameworks for relational data over the past few decades, boasting an impressive body of supporting research and methodologies~\cite{Newman2018}. However, it has become clear recently that networks in their simplest realization are incapable of 
 properly modeling 
 multi-dimensional
 relational data~\cite{Bianconi2018, Boccaletti2014, DeDomenico2015, Vasilyeva2021, 10.1088/978-0-7503-1046-8}. 
 The literature is indeed plenty with examples of
 how ignoring the multi-dimensionality of systems in network modeling leads to fundamental errors in the 
 characterization of both their structural and dynamical properties, see Refs.~\cite{radicchi2013abrupt, gomez2013diffusion, battiston_multilayer_2017, buldyrev_catastrophic_2010, Bianconi2018, Cardillo2013, DeDomenico2015, DeDomenico2015b,  de_domenico_physics_2016, diakonova_irreducibility_2016, gleeson_effects_2016, Zeng2019, Santoro2020, Osat2020, osat2017optimal} among others.

 A multiplex is probably the simplest network representation of a multi-relational system~\cite{Bianconi2018, Boccaletti2014, lee2015towards, Kivela2014, de2013mathematical}.
 A multiplex is a collection of single-layer networks sharing common nodes; each layer of a multiplex captures a different type or flavor of pairwise interaction among nodes. This is a convenient and meaningful representation for many real-world systems, including social~\cite{szell2010multirelational, mucha2010community} and biological systems~\cite{bullmore2009complex,DeDomenico2015, lim2019discordant}. 
 
Even if the system under study is truly a multiplex, data about its topology are often available in an aggregated, single-layer format. As a matter of fact, obtaining precise information about the flavor of all edges in a real multiplex could be prohibitively expensive, too time intensive, or even physically impossible. For example,  
it is relatively simple to detect correlations in coarse-level changes of gene expressions. Less simple, however, is the observation of fine-grained co-expression details, being prohibitively expensive for large genetic interaction systems. Similarly, neural connectomes 
can be reconstructed by analyzing
correlations in time series of spiking neuron activity;
however, observing the details of the interactions, i.e., synaptic junction types, is too expensive for large connectomes.
Despite the availability of data describing functional aggregate networks, the dynamics occurring on these functional multiplexes is distinct from the one happening on their observable aggregate counterparts~\cite{zanin_can_2015}.
In this respect, there is an apparent need for tools to infer full, true multiplex structure from an associated single-layer network. In this paper, we refer to this classification problem as the multiplex reconstruction problem (MRP, see Methods for its definition).

Only a few attempts to solve the MRP  exist in the literature.
In Ref.~\cite{Bagrow2021}, Bagrow and Lehmann consider the MRP in the context of temporal networks, i.e., multiplex networks where layers correspond to different snapshots of the same network at different instants of time. They obtain good prediction accuracy leveraging a sparsity-enforced lasso regression technique. The method requires full knowledge not only of the aggregated network topology, but also of the node degrees in the individual layers. 
Zhang {\it et al.} consider the MRP on two-layer multiplex networks~\cite{Zhang2021b}. Their solution of the MRP corresponds to the maximization of the edge clustering coefficient of the individual layers, obtained via a simulated-annealing-like algorithm. The algorithm is applicable only to small and sufficiently dense multiplex networks, as for example networks representing trades of different commodities among countries.
Wu {\it et al.} interpret the MRP as a statistical inference problem based on the hypothesis that network layers are instances of the configuration model~\cite{Wu2020}. They develop an expectation-maximization algorithm for solving the MRP. The algorithm can be trained on partial knowledge of the ground-truth topology of a multiplex to make predictions about the unobserved edges. 
The method allows to perform prediction and classification of edges of a multiplex network, and the classification task includes the possibility of an edge to belong simultaneously to multiple  layers. Wu {\it et al.} also provide a theoretical analysis of the MRP, relating the accuracy of reconstruction of their method to a metric of multiplex's entropy derived under the ansatz of the configuration model. 
Finally, both De Bacco {\it et al.}~\cite{DeBacco2017} and Tarres-Deulofeu {\it et al.}~\cite{tarres-deulofeu_tensorial_2019} introduce inferential methods for the analysis of multiplex networks with potentially correlated layer-wise community structure. These methods are rather general and can be used to solve various inference problems, including link prediction at the level of individual layers. Both papers consider only a 5-fold cross-validation scheme, where their classifiers are trained on 80\% of the ground-truth topology, and the 20\% of the remaining links are predicted/classified. Performance of the classifiers is excellent. Both papers study only real-world multiplex networks composed of a small number of nodes but a large number of layers, possibly because the computational time of these methods scales quadratically with the number of nodes in the system. Also, the methods may not be best suited to deal with sparse network layers, as explicitly noted by Tarres-Deulofeu {\it et al.} who state that ``the multilayer models outperform the baseline models in almost all the studied cases, except for the cases in which information is too sparse for the multilayer model to recover unobserved interactions with precision'' ~\cite{tarres-deulofeu_tensorial_2019}.

The above five articles mark, to the best of our knowledge, the only literature directly addressing the MRP. 
Some existing literature deals with the related problem of determining whether the multiplex framework is indeed required to properly model an observed single-layer network.
For example, Lacasa {\it et al.} 
show that diffusion properties of a single-layer network can be used to judge whether the network is truly represented by a single layer or if instead it is better represented by a (hidden) multiplex structure~\cite{Lacasa2018}. Similar goals are pursued by Santoro and Nicosia with an approach that leverages Kolmogorov's complexity~\cite{Santoro2020}.
Finally, Valles-Catal\'a {\it et al.} develop a method to determine the reliability of links in multiplex networks~\cite{Valles-Catala2016}. The method relies on the generalization of the stochastic block model from single-layer to multi-layer networks. Link reliability consists in determining whether an observed (or unobserved) edge exists or not. Valles-Catal\'a {\it et al.} show that accounting for the hidden multiplexity with their generalized stochastic block model indeed increases the ability to predict the existence and/or non-existence of edges. However, they do not directly focus on the MRP, i.e., the classification of observed edges in types or flavors.

 In this paper, we introduce  an  
algorithm able to approximate MRP solutions. The
computational complexity of the algorithm
grows linearly with the total number of edges in the multiplex network. 
Our algorithm is inspired by the degree-corrected stochastic block model in the sense that the probability of an observed edge to belong to a specific layer instead of another is a function of the degree of the nodes in the layers and the community structure of the layers~\cite{karrer2011stochastic}. 
The latter ingredients are learned from the partial observation of the ground-truth structure of the multiplex network.
Given that the two main ingredients of the classifier are the degree sequence and the community structure of the layers we named it as degree- and community-based classifier. 
From the systematic study of the MRP on synthetic multiplex networks, we show that the accuracy of the classification task is strongly influenced by both the heterogeneity of the individual network layers and the correlation among the structure of the layers. From the analysis of real-world networks, we show that the accuracy of the method saturates pretty quickly as the amount of  partial information available on the system increases. Roughly, having $30\%$ partial knowledge yields $90\%$ accuracy.

\section{Results}

In this paper, we consider the multiplex reconstruction problem (MRP) on multiplex networks composed of two layers only. We work under the assumption that all edges in the multiplex network are known. Depending on the experimental setup considered, we may have access to some level of ground-truth information.
We remark that in our formulation of the problem edges can belong to one layer only. The above assumptions lead to a formulation of the MRP as a binary classification task. We systematically study the MRP on both synthetic and real networks. Details on the methods, networks and experimental setups are reported in the Methods section.

\subsection{Reconstruction with degree and community information}

We begin our analysis under a peculiar experimental setup. Network layers are instances of the configuration model~\cite{molloy1995critical}. Except from the pre-imposed degree sequence, network layers are completely random so that no community structure characterizes the networks. We work with power-law degree distributions with tunable degree exponent. We assume that the degree-based (D) classifier has full knowledge of the degree sequences of both layers composing the multiplex network. Except for that, the D classifier is completely uninformed about the ground-truth topology of the system, thus the test set is composed of all edges in the network.

We use this experimental setup to understand how difficult the MRP is on synthetic multiplex networks with variable  (i) degree heterogeneity at the level of individual layers and (ii) degree-degree correlation among layers. In particular, we note that network layers are generated according to a model that is compatible with the assumption made at the basis of the D classifier, thus results from this set of experiments should represent a reliable proxy for the intrinsic difficulty of the MRP. 

\begin{figure*}[!htb]
    \centering
    \includegraphics[width=0.8\textwidth]{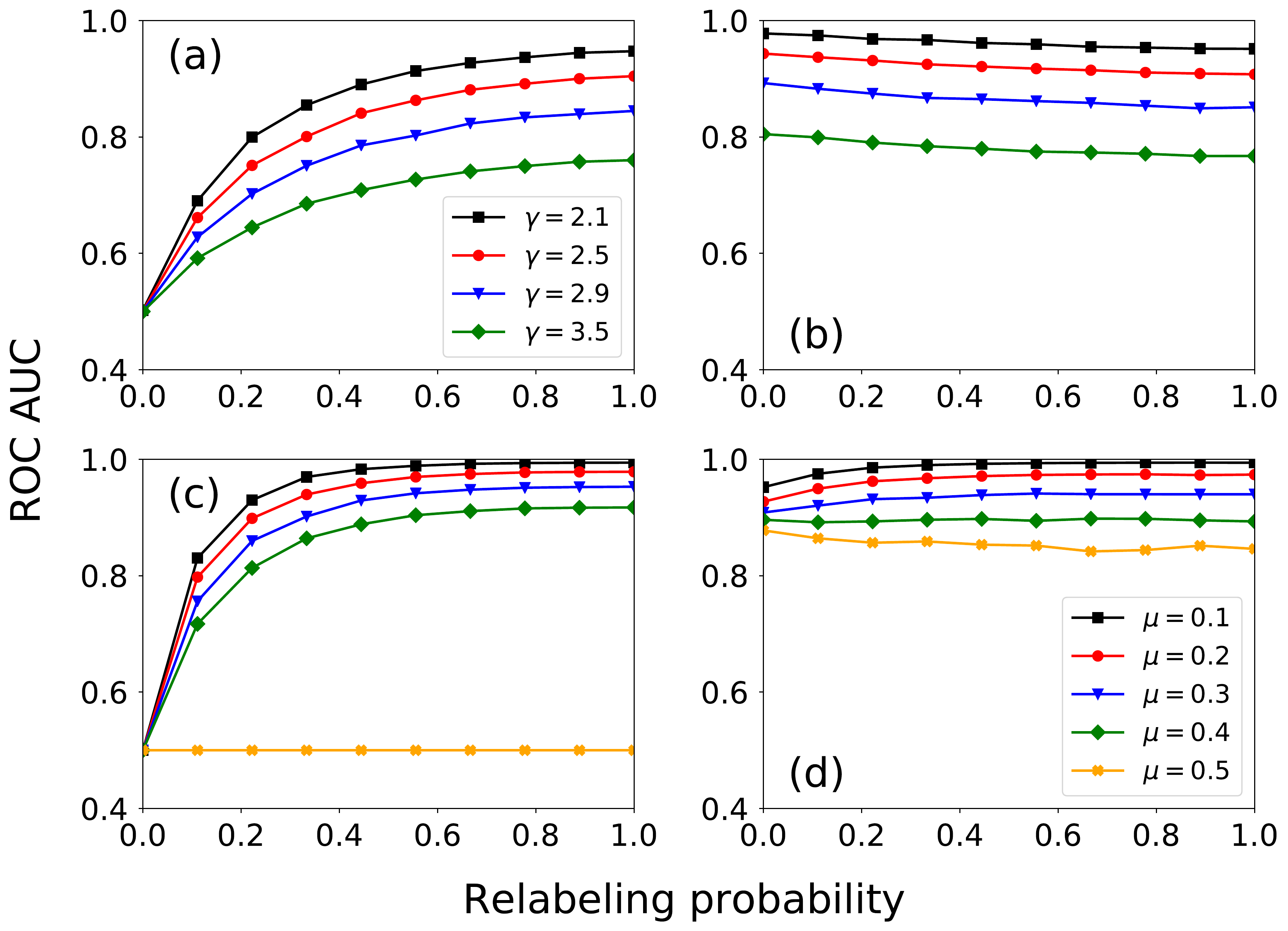}
    \caption{
        \textbf{Reconstruction of synthetic multiplex networks.}
        (a) We consider random multiplex networks with power-law degree distribution $P(k) \sim k^{-\gamma}$ composed of $N = 100,000$ nodes. Minimum degree is $k_{\min}=3$; maximum degree is $k_{\max} = \sqrt{N}$ if $\gamma \leq 3$, and $k_{\max} = N^{1/(\gamma-1)}$ if $\gamma \geq 3$. Different curves are obtained for different $\gamma$ values.  We plot the classification metric ROC AUC of the degree-based (D) classifier as a function of the probability relabeling nodes; we relabel nodes to control for degree-degree correlation (see Methods).
         Degrees are positively correlated when the relabeling probability is zero; they become progressively uncorrelated as the relabeling probability increases. 
        Results are obtained over a single realization of the model. We do not observe substantial fluctuations from run to run.
        (b) Same as in panel (a), but for multiplex networks with anti-correlated degree sequences. The sequences are maximally anti-correlated for a null relabeling probability, and they become uncorrelated as the relabeling probability increases. 
        (c) We consider the multiplex networks with pre-imposed community structure and power-law degree distribution $P(k) \sim k^{-\gamma}$. Graphs with $N = 10,000$ nodes are generated according to the Lancichinetti-Fortunato-Radicchi (LFR) model, see Methods for details. We set degree exponent $\gamma=2.1$, maximum degree $k_{\max} = \sqrt{N}$, and average degree $\langle k \rangle = 20$.  Communities have size obeying a power-law distribution with exponent $\tau = 1.0$. The strength of the community structure is determined by the mixing parameter $\mu$.
        We plot the ROC AUC of the community-based (C) classifier as a function of the probability relabeling the nodes of the multiplex. 
        The classifier is unaware of the ground-truth degree sequences of the network layers. 
        Different curves are obtained for different values of the mixing parameter $\mu$ controlling for the strength of the community structure of the network layers. The community structure of the two layers is correlated when no nodes are relabeled; correlation decreases as the relabeling probability increases. (d) We plot the ROC AUC of the degree- and community-based (DC) classifier as a function of the relabeling probability for synthetic multiplex networks with pre-imposed community structure. The networks are the same networks as in panel (c). In this case, however, the classifier takes also advantage of the knowledge of the ground-truth degree sequence of the network layers. 
    }
    \label{fig:1}
\end{figure*}

As the results of Fig.~\ref{fig:1} show, we find that the performance of our method to solve the MRP decreases as the degree exponent of the pre-imposed degree distribution increases. Essentially, solving the MRP is easier on multiplex networks with heterogeneous degree sequences than on  multiplex networks with homogeneous degree sequences. However, it is also important that the two layers are sufficiently diverse one from the other. If nodes have the same exact degrees on both layers, the performance of the D classifier is identical to that of random guessing (Fig.~\ref{fig:1}a). As the degree-degree correlation of the layers decreases, performance improves. In particular, we find that the best performance is achieved for maximally anticorrelated degree sequences (Fig.~\ref{fig:1}b). 
This is a consequence of the fact that the probability for edge to belong to a given layer is proportional to the product of the layer-wise nodes' degrees, see Eq.~(\ref{m_eq1}). When degree sequences are anti-correlated, it likely that nodes with high degree in one layer have low degree in the other layer, and edges attached to such a type of nodes are relatively easy to be correctly classified.

Second, we study the performance of the community-based (C) classifier on network layers generated according to the Lancichinetti-Fortunato-Radicchi (LFR) model~\cite{lancichinetti2008benchmark}. We use the LFR model to generate network layers with built-in community structure; we further control for the amount of degree and community-structure correlation between the layers of the multiplex. We test the performance of the classifier on these networks when informed about 
their ground-truth community structure, but no information on their degree sequence is used. As the results of Fig.~\ref{fig:1}c indicate,
the performance of the C classifier increases as the strength of the community structure increases; also, performance increases as the correlation between the community structure of the layers decreases. 
The results of Fig.~\ref{fig:1}c are due to the C classifier of Eq.~(\ref{m_eq1comm}), which expresses the probability for an edge to belong to a layer of the multiplex as proportional to the strength of the layer-wise community structure. When the community structures of the layers are anti-correlated, intra-community edges have associated a high probability to belong to the correct layer; the complementary probability is instead low due to the fact that the edge is seen as an inter-community edge in the other layer.

Finally, we apply the degree- and community-based (DC) classifier to the LFR multiplex networks. We find that the combination of the two ingredients leads to a significant boost in classification performance (Fig.~\ref{fig:1}d). Decreasing correlation leads to a visible improvement in classification performance only if  the community structure of the layers is strong enough (Fig.~\ref{fig:1}d).

\begin{figure*}[!htb]
    \centering
    \includegraphics[width=0.8\textwidth]{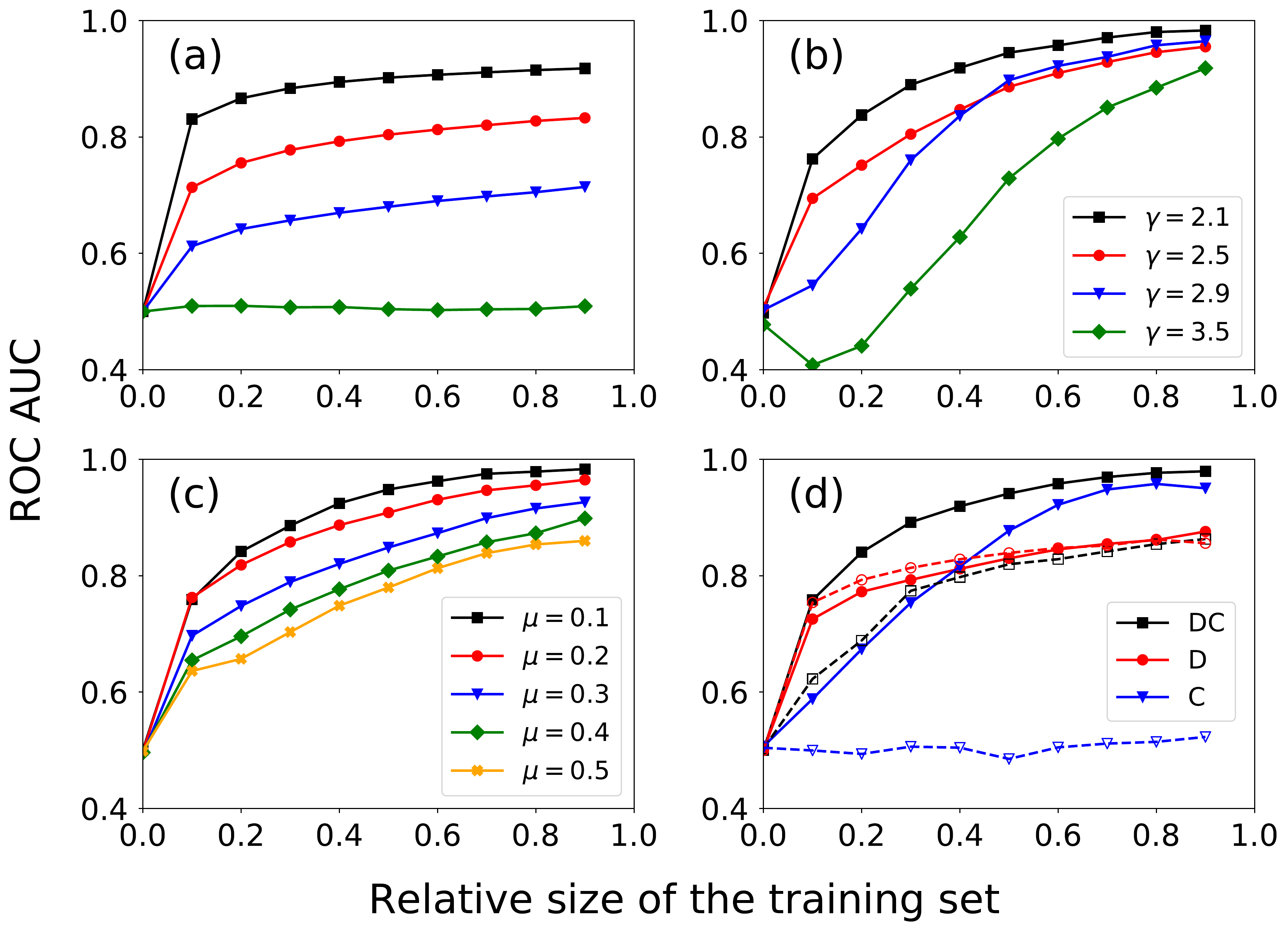}
    \caption{
    \textbf{Reconstruction of synthetic multiplex networks with partial edge information.}
    (a) We plot the performance metric ROC AUC of the degree-based (D) classifier as a function of the relative size of the training set. Multiplex networks have size $N=100,000$ and power-law degree distribution $P(k) \sim k^{-\gamma}$. We display results for different $\gamma$ values. We impose minimum degree $k_{\min} = 3$ for all $\gamma$ values, while the maximum degree is $k_{\max} = \sqrt{N}$ for $\gamma \leq 3$ and $k_{\max} = N^{1/(\gamma-1)}$ for $\gamma > 3$. The degree sequences of the layers are uncorrelated. (b) We plot the ROC AUC of the degree- and community-based (DC) classifier as a function of the relative size of the training set. Tests are performed on synthetic multiplex networks with pre-imposed community structure constructed according to the LFR model with $N = 10,000$,  $\langle k \rangle = 5.0$, and $k_{\max} = \sqrt{N}$ for $\gamma \leq 3$ and $k_{\max} = N^{1/(\gamma-1)}$ for $\gamma > 3$. Community power-law exponent is $\tau=1.0$, and mixing parameter is $\mu=0.1$. The degree sequences and community structures of the layers are uncorrelated. (c) Same as in (b), but for fixed degree exponent $\gamma = 2.1$ and different $\mu$ values. (d) We compare the performance of the different classifiers on synthetic graph with built-in community structure (filled symbols / full curves) and without community structure (empty symbols / dashed curves). We plot the performance metric of the DC classifier, the D classifier and the community-based (C) classifier. Graphs with community structure are constructed using the LFR  multiplex model with $N=10,000$, $\gamma=2.1$, $k_{\max} = \sqrt{N}$ , $\langle k \rangle = 5.0$, $\tau=1.0$, and $\mu = 0.1$. We use the configuration model with $N=10,000$, $\gamma = 2.1$, $k_{\min} = 3$, $k_{\max} = \sqrt{N}$ for the graphs with no community structure. The degree sequences and community structures of the layers are uncorrelated.
    }
    \label{fig:2}
\end{figure*}

\subsection{Reconstruction with partial edge information}

The experimental setups considered in the previous section are useful to understand intrinsic properties of the MRP. However, the setups are not very representative for applications in the real world. For example, it appears as unrealistic to have full and exact knowledge of the layer-wise community structure of the multiplex, but no information about the topology of the multiplex. From now on, we work using a standard experimental setup where we assume that the edges in the multiplex are randomly divided in a training set and a test set. The relative size of one set over the other is the main control parameter of our experiments. The classifier is trained from the knowledge of the ground-truth flavor of the edges within the training set. Performance is measured as the ROC AUC of the binary classification task concerning edges belonging to the test set. 

We consider the MRP on synthetic multiplex networks, see Fig.~\ref{fig:2}. In this set of experiments, the degree sequences and community structures of the layers are uncorrelated. Similarly to what reported in the previous section, we find that the accuracy of the DC classifier grows as the heterogeneity of the degree distribution (Fig.~\ref{fig:2}a) and the strength of the community structure (Fig.~\ref{fig:2}b and Fig.~\ref{fig:2}c) increase. 
Depending on whether the network layers have or have not assortative structure, the classification is enhanced if the classifier takes or does not take advantage of its community-based component (Fig.~\ref{fig:2}d). If the graph has no community structure, the DC classifier may also display lower performance than the simple D classifier.
However, this seems to happen only if the size of the training set is sufficiently small; further, if the system size is increased, no apparent gap in performance is longer visible, see Fig.~\ref{fig:7}. 
In all the experiments considered in Fig.~\ref{fig:2}, the accuracy of the classifier saturates quite quickly with the amount of partial information used to train the classifier. 
These general observations on the performance of the DC classifier to solve the MRP on synthetic networks are only mildly affected by the average degree of the network and its size. Specifically, we find that the performance of the classifier is almost unaffected by the average degree of multiplex network, see Fig.~\ref{fig:5}a. Also, we find that, as the system size is increased, the performance of the reconstruction algorithm mildly increases,  see Figs.~\ref{fig:5}b and~\ref{fig:5}c.

\subsection{Time complexity of the reconstruction algorithms}

\begin{figure}[!htb]
    \centering
    \includegraphics[width=0.4\textwidth]{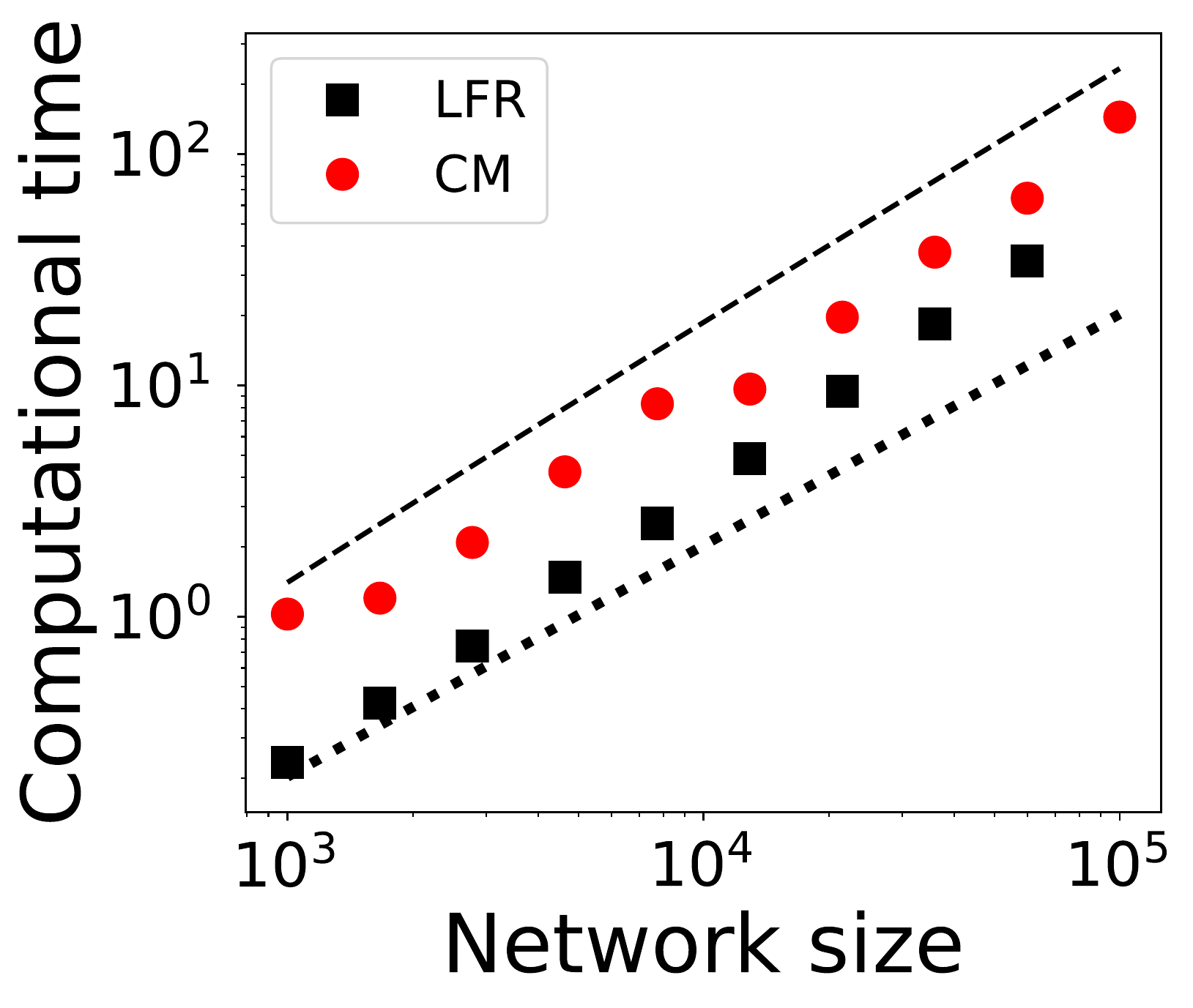}
    \caption{
    {\bf Computational complexity of algorithms for multiplex reconstruction.} We generate multiplex networks with variable size $N$. Tests are performed on synthetic multiplex networks with pre-imposed community structure constructed according to the LFR model with $\gamma=2.1$,  $\tau=1.0$, $k_{\max} = \sqrt{N}$, $\langle k \rangle = 5.0$, and $\mu =0.1$. Tests are also performed on networks without community structure that were generated using the configuration model (CM) with $\gamma =2.1$, $k_{\min}=3$ and $k_{\max} = \sqrt{N}$. No correlation at the level of degree sequence and/or community structure is present between the layers of the multiplex.
    Given a network, we provide $50\%$ of partial information to 
    the degree- and community-based (DC) classifier and measure the time required by the algorithm to reconstruct the multiplex. Computational time is measured in seconds. Simulations were run on a {\tt Intel(R) Xeon(R) CPU E5-2690 v4 @ 2.60GHz}. As guidelines, we display lines denoting the scalings $N$ (dotted) and $N \log N$ (dashed). 
        }
    \label{fig:time_complexity}
\end{figure}

The proposed DC classifier generates solutions of the MRP in a time that grows almost linearly with the total number of edges in the multiplex. The classifier benefits from the scalability of the Louvain algorithm, i.e., the method used to detect communities in the network layers~\cite{blondel2008fast}. We stress that the DC classifier can leverage any suitable community detection method to produce solutions of the MRP; however, the time complexity of the classifier may be dramatically increased by that of the community detection method. The fact that the DC method is able to reconstruct multiplex networks in linear time is verified in Fig.~\ref{fig:time_complexity}. There, we systematically apply the DC method in the reconstruction of the topology of multiplex networks composed of synthetic network layers of variable size. By construction, these networks are sparse so that the total number of edges is proportional to the total number of nodes. In the same spirit as in the analysis of Fig.~\ref{fig:2}, we use the DC classifier to generate solutions of the MRP for a given amount of partial information. Please note that, even networks that do not have pre-imposed community structure are still analyzed according to the full pipeline of the DC classifier. The quasi-linear scaling of the time of computation is apparent from our results. 

\begin{figure*}[!htb]
    \centering
    \includegraphics[width=0.8\textwidth]{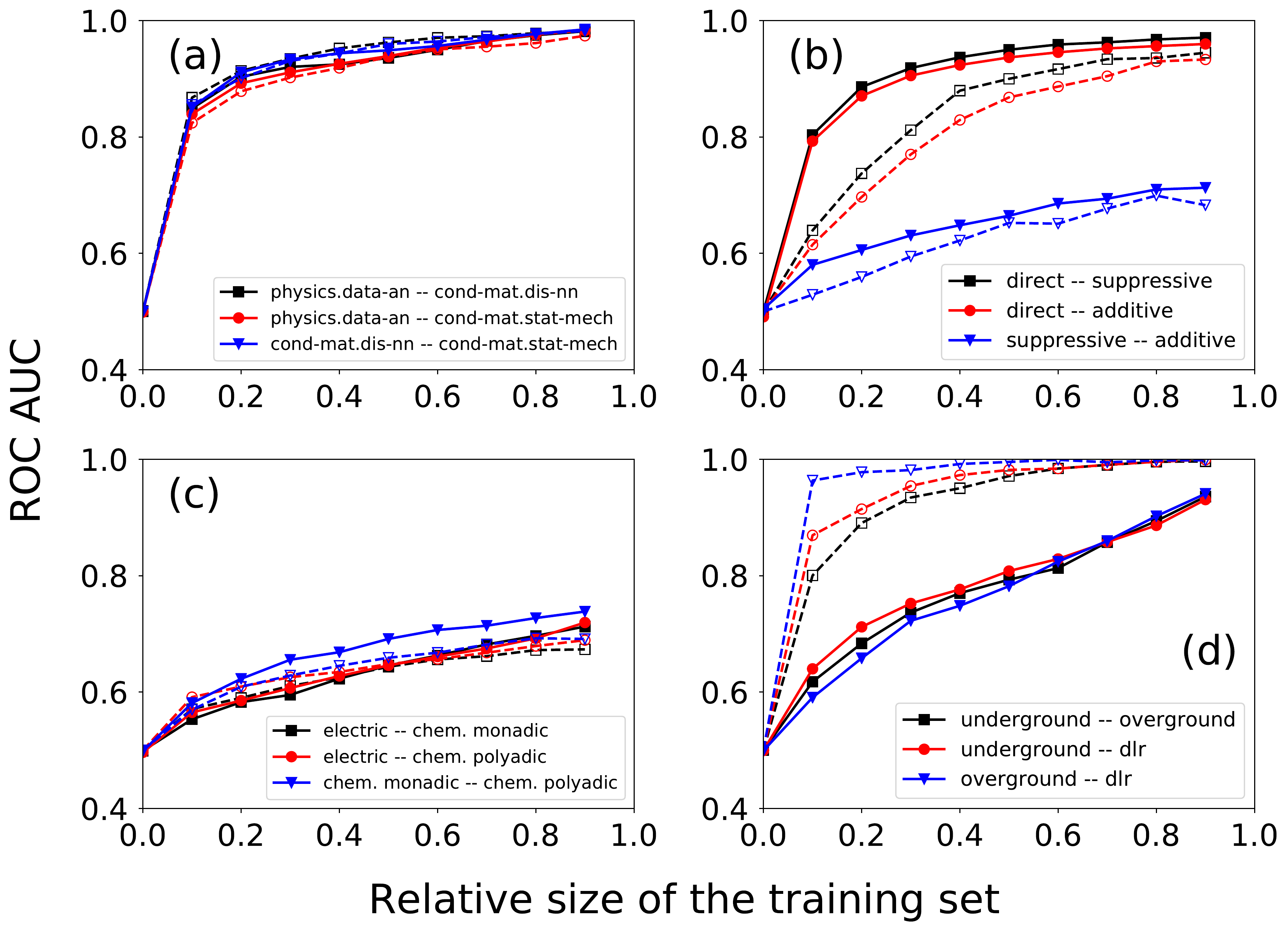}
    \caption{
    {\bf Reconstruction of real-world multiplex networks.}
    (a) ROC AUC as a function of the relative size of the training set. We report results for the arXiv  multiplex collaboration network ~\cite{de2015identifying}. 
    We compare the results achieved with the degree- and community-based (DC, filled symbols / full curves) classifier with those of the Wu {\it et al.}'s classifier (empty symbols / dashed curves)~\cite{Wu2020}. Results are averaged over $10$ realizations. (b) Same as in panel (a), but for the genetic interactions multiplex network of the {\it Drosophila Melanogaster}~\cite{de2015structural}. Results are averaged over $100$ realizations. (c) Same as in panel (a), but for the {\it Caenorhabditis Elegans} multiplex connectome~\cite{de2015muxviz}. (d) Same as in panel (a), but for the London multiplex transportation network~\cite{de2014navigability}. Results are averaged over $100$ realizations.
    }
    \label{fig:4}
\end{figure*}

\subsection{Applications to real-world multiplex networks}

Finally, we study the MRP with partial edge information on real-world multiplex networks, see Tab.~\ref{tab:summary_real_multiplexes_pairs} for the complete list of datasets analyzed. In Fig.~\ref{fig:4}, we report results for four datasets. Results for three additional datasets can be found in Fig.~\ref{fig:6}. Although some datasets include information about multiplex networks with more than two layers, we study only multiplexes formed by two layers at a time. Overall, we find a mixed behavior of the DC classifier. There are cases where the classifier is able to correctly identify the flavor of a large portion of edges even if the size of the training set is small. This fact happens for multiplex networks representing scientific collaborations (Fig.~\ref{fig:4}a) and genetic interactions in organisms (Figs.~\ref{fig:4}b and~\ref{fig:6}). In other situations instead, as for example for a connectome multiplex network (Figs.~\ref{fig:4}c) and for a multiplex transportation network (Figs.~\ref{fig:4}d), the performance of the classifier is not exceptional.

We compare the performance of the DC classifier with a suitably modified version of the classifier by Wu {\it et al.}~\cite{Wu2020}, see Methods for details. The classifier by Wu {\it et al.} relies on the configuration model, and the classification of edges is obtained via the expectation-maximization algorithm.  No information on the network community structure is used by the method. The classifier by Wu {\it et al.} excels on the London multiplex transportation network (Fig.~\ref{fig:4}d). It produces comparable performance to the one of the DC classifier in all other multiplex networks, except for those characterized by layers with sufficiently strong but uncorrelated community structure, as for example the genetic iteraction multiplex of the {\it Drosophila Melanogaster} (Fig.~\ref{fig:4}a). The strength of the community structure of a network layer is measured in terms of modularity, see Tab.~\ref{tab:summary_real_multiplexes_pairs} for modularity values of the best partitions for the various multiplex networks. In the same table, we also report the normalized mutual information between the best partitions of the two layers, which provides us with a quantitative proxy to judge the level of similarity between pairwise community structures.

In Fig.~\ref{fig:8}, we replicate the analysis for real networks by finding communities with Infomap~\cite{rosvall2008maps} rather than Louvain. Although the two community detection algorithms typically find different partitions of the network layers, the performance of the DC classifier is similar in the two cases, indicating that the use of assortative structure, irrespective of its details, enhances the reconstruction of a multiplex ground-truth topology.

\section{Discussion}

 In this paper, we studied the multiplex reconstruction problem (MRP), i.e., the identification of the flavour of edges in multiplex networks composed of two layers. 
 
 First, we characterized facets of the reconstruction problem.  We showed that the intrinsic difficulty of the MRP depends on the level of heterogeneity of the individual layers, in the sense that multiplex networks with broad degree distributions are easier to reconstruct than multiplex networks with homogeneous degree distributions. If layers are characterized by community structure, then the stronger such a structure is the easier the MRP is. However, within-layer structural diversity is not the only important ingredient that determines the hardness of the MRP. The type and strength of structural correlation among layers is essential too. We showed that multiplex networks that have non-correlated or anticorrelated layers, at the level of degree sequences and/or community structure, can be reconstructed quite well. On the other hand, positively correlated layers do not allow for an easy reconstruction.
 
 Second, we presented a new algorithm for solving the MRP. 
The method is inspired by the degree-corrected stochastic block model, in the sense that the probability of two nodes to be connected is proportional to the product of their degrees and a factor that accounts for the community structure of the network.
 In our experiments, the classifier is trained on a portion of edges whose ground-truth layer is revealed. Its performance is measured in terms of the ability of classification of the remaining edges whose ground-truth layer is not revealed. We systematically studied the performance of the classifier on both synthetic and real multiplex networks as a function of the relative size of the training set of edges. We found that the performance of the classifier saturates quite quickly as the amount of information used to train the classifier increases. Roughly, $30\%$ of known edges are sufficient to let the classifier reach $90\%$ performance. There are, however, also multiplex networks that are not easily reconstructed by our method.
 
Our main results are based on community partitions identified via the modularity maximization algorithm Louvain. However, we showed that comparable performance can be achieved using Infomap, which determines community structure solving a different optimization problem. This means that assortative structure learned from partial observation is useful for the reconstruction task irrespective of its details.

 The present work can be extended in multiple directions. The generalization of the MRP to multiplex networks composed of more than two layers is one of these directions. Considering mechanisms of aggregation different from the exclusive OR that we addressed in this paper is another potentially relevant direction. Also, the two main ingredients of the reconstruction algorithm developed in this paper can be used in more sophisticated, maybe more effective ways. For example, one could think of treating the MRP as a maximum likelihood problem where the network layers of the multiplex are fitted by degree-corrected stochastic block models, and edges's flavors are treated as the tunable parameters of the fit. Such an extension could also involve the use of more sophisticated models such as the multi-layer degree-corrected stochastic block model~\cite{Valles-Catala2016}. Potential solutions of the MRP could be then obtained by means of likelihood maximization using standard optimization techniques such as the expectation-maximization algorithm or simulated annealing.

 \acknowledgments{
S.P. and F.R. acknowledge support by the Army Research Office (W911NF-21-1-0194); F.R.  acknowledges support by the Air Force Office of Scientific Research (FA9550-21-1-0446). The funders had no role in study design, data collection and analysis, decision to publish, or  any opinions, findings, and conclusions or recommendations expressed in the manuscript.}

\section{Methods}

\subsection{The multiplex reconstruction problem}

We consider multiplex networks composed of two layers, namely $\alpha$ and $\beta$. The graph representing layer $\alpha$ is denoted by $\mathcal{G}^{(\alpha)} = \{ \mathcal{N}^{(\alpha)},  \mathcal{E}^{(\alpha)}\}$, where $\mathcal{N}^{(\alpha)}$ is the set of nodes of the layer, and $\mathcal{E}^{(\alpha)}$ is the set of its edges. The same notation is used for layer $\beta$. 
Being a multiplex, we assume that $\mathcal{N}^{(\alpha)} = \mathcal{N}^{(\beta)}$, and we indicate the size of the network with $N = \left|\mathcal{N}^{(\alpha)}\right| = \left|\mathcal{N}^{(\beta)}\right|$. Further, we exclude the possibility that the edge $(i,j)$ belongs simultaneously to both layers, meaning that  $\mathcal{E}^{(\alpha)} \cap \mathcal{E}^{(\beta)} = \emptyset$. 
This choice is mainly dictated by simplicity. We stress, however, that ours is a quite reasonable assumption. All synthetic networks considered in our analysis are indeed characterized by a negligible number of edges shared by the layers. The same observation can be made also for several real-world networks studied in this paper, see Table~\ref{tab:summary_real_multiplexes_pairs}.

 We indicate the degree sequence of the layer $\alpha$ with $\vec{k}^{(\alpha)} = (k_1^{(\alpha)}, k_2^{(\alpha)}, \ldots, k_N^{(\alpha)})$, where $k_i^{(\alpha)}$ is the degree of node $i$ in layer $\alpha$ defined as
 
 \begin{equation} \label{eq:degree}
    k^{(\alpha)}_i = \sum_{(r,s) \in \mathcal{E}^{(\alpha)}} \, \delta(i,r)+\delta(i,s) \, .
\end{equation}
In the above expression, $\delta(x,y)$ is the Kronecker delta function, i.e., $\delta(x,y) = 1$ if $x=y$ and $\delta(x,y) =0 $ if $x \neq y$. A similar expression is used to relate the degree sequence $\vec{k}^{(\beta)}$ to the  set $\mathcal{E}^{(\beta)}$.

 The community memberships of the nodes in the two layers are denoted respectively as  $\vec{\sigma}^{(\alpha)}$ and $\vec{\sigma}^{(\beta)}$. Community structure is either {\it a priori} known based on the generative network model of the multiplex layers, or inferred by the Louvain~\cite{blondel2008fast} or the Infomap~\cite{rosvall2008maps}) algorithm. Potentially, other community detection algorithms can be used as well.

We define the multiplex reconstruction problem (MRP) as the binary classification of the individual edges in the multiplex layers, i.e.,  predicting whether the generic edge $(i,j)$ belongs to either  $\mathcal{E}^{(\alpha)}$ or $\mathcal{E}^{(\beta)}$. 

We perform the classification using partial knowledge of the ground-truth multiplex network. We consider two  experimental setups: 

\begin{enumerate}
    \item We assume to have full knowledge about some of the features of the individual nodes, i.e., degrees and/or community memberships, but no direct information about the sets $\mathcal{E}^{(\alpha)}$ and $\mathcal{E}^{(\beta)}$. We train a classifier using the available information and apply it to the classification of all edges $(i,j) \in  \mathcal{E}^{(\alpha)} \cup \mathcal{E}^{(\beta)}$.

    \item We assume to have information about a portion of the network edges, namely the training sets $\mathcal{E}_{\textrm{train}}^{(\alpha)} \subseteq \mathcal{E}^{(\alpha)}$ and $\mathcal{E}_{\textrm{train}}^{(\beta)} \subseteq \mathcal{E}^{(\beta)}$, respectively. We train a classifier on these sets and we use it to classify all edges $(i,j) \in  \mathcal{E}_{\textrm{test}}^{(\alpha)} \cup \mathcal{E}_{\textrm{test}}^{(\beta)}$, with
    $\mathcal{E}_{\textrm{test}}^{(\alpha)} =\mathcal{E}^{(\alpha)} \setminus \mathcal{E}_{\textrm{train}}^{(\alpha)}$ and 
    $\mathcal{E}_{\textrm{test}}^{(\beta)} =\mathcal{E}^{(\beta)} \setminus \mathcal{E}_{\textrm{train}}^{(\beta)}$.
\end{enumerate}

In both the above experimental setups, we measure the performance of the classifier using the area under the receiver operating characteristic curve (ROC AUC), i.e., a standard metric in binary classification tasks.

\subsection{Degree-based classifier}

In the first experimental setup, we assume to have complete knowledge of the degree sequences $\vec{k}^{(\alpha)}$ and $\vec{k}^{(\beta)}$. 
We use a straightforward prediction model, where we pretend that the networks of layers $\alpha$ and $\beta$ are random networks generated according to the configuration model with prescribed degree sequences $\vec{k}^{(\alpha)}$ and $\vec{k}^{(\alpha)}$, respectively~\cite{molloy1995critical}. Note that the network topology of each layer is assumed to be generated independently. According to our classifier, the edge $(i, j) \in \mathcal{E}^{(\alpha)} \cup \mathcal{E}^{(\beta)}$ belongs to layer $\alpha$ with probability

\begin{equation} \label{m_eq1}
    P\left[ (i,j) \in \mathcal{E}^{(\alpha)} | \vec{k}^{(\alpha)}, \vec{k}^{(\beta)} \right] = \frac{k_i^{(\alpha)} k_j^{(\alpha)} } { k_i^{(\alpha)} k_j^{(\alpha)} + k_i^{(\beta)} k_j^{(\beta)} } \; .
\end{equation}

Clearly, we have that 
\begin{align}
    & P\left[ (i,j) \in \mathcal{E}^{(\beta)} | \vec{k}^{(\alpha)}, \vec{k}^{(\beta)} \right] = \nonumber\\
    & \quad\qquad 1 - P\left[ (i,j) \in \mathcal{E}^{(\alpha)} | \vec{k}^{(\alpha)}, \vec{k}^{(\beta)} \right].
\end{align}

In the second experimental setup, we first use Eq.~(\ref{eq:degree}) to estimate the degree sequences $\vec{k}^{(\alpha)}_{\textrm{train}}$ and $\vec{k}^{(\beta)}_{\textrm{train}}$ from the 
the sets $(\mathcal{E}^{(\alpha)} \cup \mathcal{E}^{(\beta)})\setminus\mathcal{E}_{\textrm{train}}^{(\beta)}$ and $(\mathcal{E}^{(\alpha)} \cup \mathcal{E}^{(\beta)})\setminus\mathcal{E}_{\textrm{train}}^{(\alpha)}$, respectively. Essentially, we pretend that a layer  is formed by all edges in the aggregate less those edges that we know for sure they belong to the other layer.
Then, we 
apply the  classifier of Eq.~(\ref{m_eq1}), where  $\vec{k}^{(\alpha)}_{\textrm{train}}$ and $\vec{k}^{(\beta)}_{\textrm{train}}$ respectively replace the unknown ground-truth vectors $\vec{k}^{(\alpha)}$ and $\vec{k}^{(\beta)}$, to the edges in the test set $\mathcal{E}_{\textrm{test}}^{(\alpha)} \cup \mathcal{E}_{\textrm{test}}^{(\beta)}$.

\subsection{Community-based classifier}

In the first experimental setup, we assume to have complete knowledge of the community memberships  $\vec{\sigma}^{(\alpha)}$ and $\vec{\sigma}^{(\beta)}$. 
We use a straightforward prediction model, where we pretend that the networks of layers $\alpha$ and $\beta$ are random networks generated according to the homogeneous stochastic block model with prescribed community memberships $\vec{\sigma}^{(\alpha)}$ and $\vec{\sigma}^{(\alpha)}$, respectively. Note that the network topology of each layer is assumed to be generated independently. According to our classifier, the edge $(i, j) \in \mathcal{E}^{(\alpha)} \cup \mathcal{E}^{(\beta)}$ belongs to layer $\alpha$ with probability

\begin{equation} \label{m_eq1comm}
    P\left[ (i,j) \in \mathcal{E}^{(\alpha)} | \vec{\sigma}^{(\alpha)}, \vec{\sigma}^{(\beta)} \right] = \frac{C(\sigma^{(\alpha)}_i,\sigma^{(\alpha)}_j)}{C(\sigma^{(\alpha)}_i,\sigma^{(\alpha)}_j)+C(\sigma^{(\beta)}_i,\sigma^{(\beta)}_j)} \; .
\end{equation}

Clearly, we have that 
\begin{align}
    & P\left[ (i,j) \in \mathcal{E}^{(\beta)} | \vec{\sigma}^{(\alpha)}, \vec{\sigma}^{(\beta)} \right] = \nonumber\\
    & \quad\qquad 1 - P\left[ (i,j) \in \mathcal{E}^{(\alpha)} | \vec{\sigma}^{(\alpha)}, \vec{\sigma}^{(\beta)} \right].
\end{align}

In Eq.~(\ref{m_eq1comm}), $C(\sigma_i,\sigma_j)$ 
 represents the propensity
 that two nodes with community memberships $\sigma_i$ and $\sigma_j$ are connected. For simplicity, we assume that such a 
 propensity
 can be written as 
\begin{equation} \label{eq_comm}
C(\sigma_i,\sigma_j) = \nu \,  [1 - \delta(\sigma_i, \sigma_j)] + (1-\nu) \, \delta(\sigma_i, \sigma_j)
\; ,
\end{equation}
with $0 \leq \nu \leq 1$. Please note that the value of the parameter $\nu$ is known to the classifier.

The assumption of Eq.~(\ref{eq_comm}) is similar to the one underlying the Lancichinetti-Fortunato-Radicchi (LFR)  model~\cite{lancichinetti2008benchmark}, with the caveat $\nu \simeq \mu$, where $\mu$ is the mixing parameter of the LFR model.

In the second experimental setup, we 
apply a community detection algorithm to
the network formed by the edges in the set $(\mathcal{E}^{(\alpha)} \cup \mathcal{E}^{(\beta)})\setminus\mathcal{E}_{\textrm{train}}^{(\beta)}$ to infer the community structure $\vec{\sigma}^{(\alpha)}_{\textrm{train}}$ of layer $\alpha$. A similar procedure is used to infer the vector $\vec{\sigma}^{(\beta)}_{\textrm{train}}$.
Finally, 
we use the classifier of Eq.~(\ref{m_eq1comm}), where  $\vec{\sigma}^{(\alpha)}_{\textrm{train}}$ and $\vec{\sigma}^{(\beta)}_{\textrm{train}}$ are used in place of 
$\vec{\sigma}^{(\alpha)}$ and $\vec{\sigma}^{(\beta)}$, respectively. The parameter $\nu$ of Eq.~(\ref{eq_comm}) is 
estimated as
\begin{equation} \label{eq_mu}
        \nu_{\textrm{train}} = 1 -  \frac{R_{\textrm{train}}^{(\alpha)}  +   R_{\textrm{train}}^{(\beta)} } 
        {\left|(\mathcal{E}^{(\alpha)} \cup \mathcal{E}^{(\beta)})\setminus\mathcal{E}_{\textrm{train}}^{(\beta)}\right| + \left| (\mathcal{E}^{(\alpha)} \cup \mathcal{E}^{(\beta)})\setminus\mathcal{E}_{\textrm{train}}^{(\alpha)}\right|}
        \; ,
\end{equation}    
where
\[
R_{\textrm{train}}^{(\alpha)} = 
\sum_{(i,j) \in (\mathcal{E}^{(\alpha)} \cup \mathcal{E}^{(\beta)})\setminus\mathcal{E}_{\textrm{train}}^{(\beta)}}
\delta\left[ (\vec{\sigma}^{(\alpha)}_{\textrm{train}})_i , (\vec{\sigma}^{(\alpha)}_{\textrm{train}})_j \right] \;.
\]
A similar definition is used for $R_{\textrm{train}}^{(\beta)}$.

It is important to note that the goal of the classifier is not to infer meaningful community structure in the layers of the multiplex. Rather, we are solely concerned with leveraging assortative group structure in the reconstruction process. Therefore, the above choice of using Louvain (or Infomap) is a pragmatic one; Louvain (or Infomap) is a fast algorithm to discover assortative communities. If the communities found are spurious, then this fact will be accounted for in the parameter learned in Eq.~(\ref{eq_mu}), which in turn will generate scores in Eq.~(\ref{eq_comm}) that are similar for intra- and inter-community edges.

\subsection{Degree- and community-based classifier}

In the first experimental setup, we 
assume to have complete knowledge of the degree sequences $\vec{k}^{(\alpha)}$ and $\vec{k}^{(\beta)}$ of the network layers, and of the community memberships $\vec{\sigma}^{(\alpha)}$ and $\vec{\sigma}^{(\beta)}$. 
We define a score that is inspired by the edge connection probability in the degree-corrected stochastic block model~\cite{karrer2011stochastic}.
Also in this case, we assume that the network topology of each layer is generated independently of the other. According to this degree- and community-based classifier, the edge $(i, j) \in \mathcal{E}^{(\alpha)} \cup \mathcal{E}^{(\beta)}$ belongs to layer $\alpha$ with probability

\begin{align}\label{m_eq2}
    & 
        P\left[ (i,j) \in \mathcal{E}^{(\alpha)} | \vec{k}^{(\alpha)}, \vec{k}^{(\beta)}, \vec{\sigma}^{(\alpha)}, \vec{\sigma}^{(\beta)} \right]  
        = \nonumber\\
    & \quad\qquad Q \,  
         k_i^{(\alpha)} k_j^{(\alpha)} \, C(\sigma^{(\alpha)}_{i}, \sigma^{(\alpha)}_{j})
            \; .
\end{align}
In the above equation, $Q$ is a normalization constant so that
\begin{align}
    & P\left[ (i,j) \in \mathcal{E}^{(\beta)} | \vec{k}^{(\alpha)}, \vec{k}^{(\beta)}, \vec{\sigma}^{(\alpha)}, \vec{\sigma}^{(\beta)} \right] = \nonumber\\
    & \qquad 1 - P\left[ (i,j) \in \mathcal{E}^{(\alpha)} | \vec{k}^{(\alpha)}, \vec{k}^{(\beta)}, \vec{\sigma}^{(\alpha)}, \vec{\sigma}^{(\beta)} \right] \; .
\end{align}
 In Eq.~(\ref{m_eq2}), $C(\sigma_i,\sigma_j)$ 
 represents the propensity
 that two nodes with community memberships $\sigma_i$ and $\sigma_j$ are connected. 
  For simplicity, we assume that such a 
 propensity
 can be written as in Eq.~(\ref{eq_comm}), and that the value of the parameter $\mu$ is known.
 
In the second experimental setup, 
we pretend that layer $\alpha$ is formed by all edges in the set $(\mathcal{E}^{(\alpha)} \cup \mathcal{E}^{(\beta)})\setminus\mathcal{E}_{\textrm{train}}^{(\beta)}$, and layer $\beta$ is composed of all edges in the set $(\mathcal{E}^{(\alpha)} \cup \mathcal{E}^{(\beta)})\setminus\mathcal{E}_{\textrm{train}}^{(\alpha)}$.
We then use Eq.~(\ref{eq:degree}) to estimate the degree sequences $\vec{k}^{(\alpha)}_{\textrm{train}}$ and $\vec{k}^{(\beta)}_{\textrm{train}}$, and a
community detection algorithm to infer the layer community structures $\vec{\sigma}^{(\alpha)}_{\textrm{train}}$ and $\vec{\sigma}^{(\beta)}_{\textrm{train}}$.
Finally, 
we use the classifier of Eq.~(\ref{m_eq2}), where  $\vec{k}^{(\alpha)}_{\textrm{train}}$ and $\vec{k}^{(\beta)}_{\textrm{train}}$ respectively replace the unknown ground-truth vectors $\vec{k}^{(\alpha)}$ and $\vec{k}^{(\beta)}$, and $\vec{\sigma}^{(\alpha)}_{\textrm{train}}$ and $\vec{\sigma}^{(\beta)}_{\textrm{train}}$ are used in place of 
$\vec{\sigma}^{(\alpha)}$ and $\vec{\sigma}^{(\beta)}$, respectively.
The parameter $\nu_{\textrm{train}}$ of Eq.~(\ref{eq_comm}) is  estimated with Eq.~(\ref{eq_mu}).

\subsection{Modified Wu {\it et al.} classifier}

The original algorithm by Wu {\it et al.} is conceived to solve a more complicated problem than the one considered here~\cite{Wu2020}. Specifically, it aims at predicting the existence of an edge, and if the edge exists, it aims at classifying the flavor of the edge. In the classification part of the problem, an edge can belong to one layer, the other, or both of them. 

We modify the Wu {\it et al.} algorithm to solve the classification problem considered in this paper. We use this classifier only in the second experimental setup, where partial information on the multiplex edges is provided with the training sets $\mathcal{E}_{\textrm{train}}^{(\alpha)}$ and $\mathcal{E}_{\textrm{train}}^{(\beta)}$. As in its original formulation, also here the classifier takes advantage of the expectation-maximization (EM) algorithm.
First, we define the known degree of node $i$ in layer $\alpha$ as
\begin{equation}
z^{(\alpha)}_i = \sum_{(r,s) \in \mathcal{E}^{(\alpha)}_{\textrm{train}} } \,  \delta(i,r)  + \delta(i,s)  \; .
\label{eq:wu1}
\end{equation}
A similar equation is used to define $z^{(\beta)}_i$. Then, we initialize $P[(i,j) \in  \mathcal{E}^{(\alpha)}_{\textrm{test}} | \vec{\kappa}^{(\alpha)}, \vec{\kappa}^{(\beta)}] = P[(i,j) \in  \mathcal{E}^{(\beta)}_{\textrm{test}} | \vec{\kappa}^{(\alpha)}, \vec{\kappa}^{(\beta)}] = 1/2$ for all edges in the test set. We then iterate the following equations:

\begin{equation}
\begin{array}{l}
\kappa^{(\alpha)}_i = 
z^{(\alpha)}_i +  
\\
\sum_{(r,s) \in \mathcal{E}^{(\alpha)}_{\textrm{test}} } \, \left[ \delta(i,r)  + \delta(i,s) \right] \, P[(r,s) \in  \mathcal{E}^{(\alpha)}_{\textrm{test}} | \vec{\kappa}^{(\alpha)}, \vec{\kappa}^{(\beta)}]
\end{array}
\label{eq:wu2}
\end{equation}
and
\begin{equation}
P[(i,j) \in  \mathcal{E}^{(\alpha)}_{\textrm{test}} | \vec{\kappa}^{(\alpha)}, \vec{\kappa}^{(\beta)}] = \frac{\kappa^{(\alpha)}_i \, \kappa^{(\alpha)}_j}{\kappa^{(\alpha)}_i \, \kappa^{(\alpha)}_j + \kappa^{(\beta)}_i \, \kappa^{(\beta)}_j} \; .
\label{eq:wu3}
\end{equation}
We use a similar expression as of Eq.(\ref{eq:wu2}) for $\kappa^{(\beta)}_i$. Note that $P[(i,j) \in  \mathcal{E}^{(\alpha)}_{\textrm{test}} | \vec{\kappa}^{(\alpha)}, \vec{\kappa}^{(\beta)}] = 1 - P[(i,j) \in  \mathcal{E}^{(\beta)}_{\textrm{test}} | \vec{\kappa}^{(\alpha)}, \vec{\kappa}^{(\beta)}]$. At the end of each iteration, we check for convergence by comparing the predicted values of the probabilities $P[(i,j) \in  \mathcal{E}^{(\alpha)}_{\textrm{test}} | \vec{\kappa}^{(\alpha)}, \vec{\kappa}^{(\beta)}]$ with those predicted in the previous iteration of the algorithm. In our tests, the condition for convergence is satisfied if the difference between consecutive values of the estimated probabilities is smaller than $\epsilon = 10^{-3}$ for all edges in the test set.   

\subsection{Multiplex networks}
\subsubsection*{Synthetic multiplex networks}

We begin by generating a degree sequence from the power-law distribution $P(k) \sim k^{-\gamma}$ for $k \in [3, k_{\max}]$, and $P(k) =0$ otherwise. As prescribed in Ref.~\cite{catanzaro2005generation}, we set $k_{\max} = \sqrt{N}$ if $2 < \gamma \leq 3$, and $k_{\max} = N^{1/(\gamma-1)}$ for $\gamma > 3$. We consider various values of the degree exponent $\gamma$.

To generate positively correlated degree sequences, we sort the obtained degree sequence and use it for layer $\alpha$, i.e., $k^{(\alpha)}_1 \leq k^{(\alpha)}_2 \leq \ldots \leq k^{(\alpha)}_N$. The degree sequence of layer $\beta$ is a copy of the one of layer $\alpha$, i.e., $k^{(\beta)}_i = k^{(\alpha)}_i$ for all $i =1, \ldots, N$. This condition creates perfectly correlated degree sequences. To decrease correlation between the degree sequences of the layers, we swap the labels of random pairs of nodes in layer $\beta$. The level of correlation is dependent on the fraction of pairs that undergo swapping. If enough pairs of node labels are swapped, the two degree sequences are completely uncorrelated. The resulting degree sequences are then independently used to generate the network layers of the multiplex using the configuration model~\cite{molloy1995critical}.

To obtain anticorrelated degree sequences, we start from a sorted degree sequence for layer $\alpha$. For layer $\beta$, we  use $k^{(\beta)}_i = k^{(\alpha)}_{N-i}$ for all $i = 1, \ldots, N$. This fact ensures that the two degree sequences are maximally anticorrelated.
Also here, we swap the labels of a certain fraction of pairs of nodes in layer $\beta$ to decrease the level of correlation between the two degree sequences. The resulting degree sequences are then independently used to generate the network layers of the multiplex using the configuration model~\cite{molloy1995critical}.

To study the effect of correlation between the layerwise community structures
on the multiplex reconstruction problem
we generate network layers using the LFR model~\cite{lancichinetti2008benchmark}.
We begin by generating an instance of the LFR model with given set of parameters. We fix the value of the community size power-law exponent $\tau = 1.0$. We consider various values of the  average degree $\langle k \rangle$.
Also, we vary the degree exponent $\gamma$. We do not impose any constraint on the size and number of communities. We consider various values of the mixing parameter $\mu$. The same network instance is used as topology for both network layers $\alpha$ and $\beta$. 
To reduce the edge overlap between the layers to a negligible value without altering the correlation of the layers' community structure, we swap at random the labels of all pairs of nodes within the same community in layer $\beta$.
To reduce the correlation among the structure of the network layers, we swap the labels for a random fraction of pairs of nodes in layer $\beta$.

\subsubsection*{Real multiplex networks}

We analyze several real-world multiplex networks. Even if some dataset consists of more than two layers, our tests are performed considering two layers at a time. For a given combination of the layers, edges shared by both layers are deleted and no information on their existence is considered in the MRP. 
For completeness, we report the number of edges shared by the layers of a multiplex network, namely $|\mathcal{E}^{(\alpha, \beta)}|$ in Table~\ref{tab:summary_real_multiplexes_pairs}. These generally account for a very small number of edges compared to the total number of edges that are not shared by the layers. Specifically, the ratio $R_{\textrm{shared}} = |\mathcal{E}^{(\alpha, \beta)}| / ( |\mathcal{E}^{(\alpha)}| + |\mathcal{E}^{(\beta)}| ) $ is always smaller than $0.150$ except for two cases: (i) the layers ``suppressive'' and ``additive'' for the {\it Drosophila Melanogaster} multiplex for which $R_{\textrm{shared}} = 0.166$, and (ii) the layers ``chem. monadic'' and ``chem. polyadic'' of the {\it Caenorhabditis Elegans} multiplex for which $R_{\textrm{shared}} = 0.321$.

Then, the set of nodes in the corresponding multiplex is given by the union of the sets of nodes of the two individual layers. Details on the datasets analyzed in the paper are reported in Table~\ref{tab:summary_real_multiplexes_pairs}. In the table, we report also values of the modularity value for the best partition detected by the Louvain algorithm on each layer. We further measure the normalized mutual information between the layer-wise partitions to assess their similarity.

\begin{table*}[!htb]
    \centering
    \begin{tabular}{|l||l|l|r|r|r|r|r|r|r|}
        \hline
        Dataset & $\alpha$ & $\beta$ & $N$ & $|\mathcal{E}^{(\alpha)}|$ &  $|\mathcal{E}^{(\beta)}|$ & $|\mathcal{E}^{(\alpha, \beta)}|$  & $Q^{(\alpha)}$ & $Q^{(\beta)}$ & NMI \\
        \hline
        \hline
         & physics.data-an & cond-mat.dis-nn & $7,187$ & $11,929$ & $4,785$ & $2,556$ 
         & $0.89$ & $0.96$ & $0.77$ \\
         arXiv~\cite{de2015identifying} & physics.data-an & cond-mat.stat-mech & $5,963$ & $13,326$ & $1,423$ 
         & $1,159$ 
         & $0.86$ & $0.98$ & $0.75$ \\
        & cond-mat.dis-nn & cond-mat.stat-mech & $4,342$ & $6,839$ & $2,080$ 
        & $502$ 
        & $0.91$ & $0.97$ & $0.78$ \\
        \hline
        & direct & suppressive & $7,519$ & $23,911$ & $1,798$ 
        & $66$ 
        & $0.45$ & $0.64$ & $0.49$ \\
         D. Melanogaster~\cite{de2015structural} & direct & additive & $7,486$ & $23,928$ & $1,376$ 
         & $49$ 
         & $0.46$ & $0.67$ & $0.48$ \\
        & suppressive & additive & $1,005$ & $1,395$ & $956$ 
        & $469$ 
        & $0.66$ & $0.73$ & $0.59$ \\
        \hline
        & electric & chem. monadic & $273$ & $406$ & $777$ 
        & $111$ 
        & $0.67$ & $0.51$ & $0.37$ \\
         C. Elegans~\cite{de2015muxviz} & electric & chem. polyadic & $277$ & $355$ & $1,541$ 
         & $162$ 
         & $0.70$ & $0.44$ & $0.38$ \\
        & chem. monadic & chem. polyadic & $273$ & $258$ & $1,073$ 
        & $630$ 
        & $0.70$ & $0.40$ & $0.37$ \\
        \hline
        & underground & overground & $321$ & $301$ & $72$ 
        & $11$ 
        & $0.82$ & $0.78$ & $0.67$ \\
         London~\cite{de2014navigability} & underground & dlr & $311$ & $312$ & $46$ 
         & $0$ 
         & $0.83$ & $0.70$ & $0.69$ \\
        & overground & dlr & $126$ & $83$ & $46$ 
        & $0$ 
        & $0.77$ & $0.69$ & $0.70$ \\
        \hline
        & phys. assoc. & dir. iteract. & $2,577$ & $1,038$ & $6,622$ 
        & $445$ 
        & $0.87$ & $0.51$ & $0.51$ \\
         S. Pombe~\cite{de2015structural} & phys. assoc. & colocalization & $3,009$ & $1,431$ & $30,752$ 
         & $52$ 
         & $0.84$ & $0.23$ & $0.44$ \\
        & dir. iteract. & colocalization & $3,782$ & $6,940$ & $30,677$ 
        & $127$ 
        & $0.52$ & $0.23$ & $0.40$ \\
        \hline
        & phys. assoc. & dir. iteract. & $17,415$ & $37,361$ & $72,004$ 
        & $12,284$ 
        & $0.53$ & $0.41$ & $0.32$ \\
         H. Sapiens~\cite{de2015structural} & phys. assoc. & colocalization & $15,254$ & $82,050$ & $16,189$ 
         & $1,386$ 
         & $0.42$ & $0.64$ & $0.35$ \\
        & dir. iteract. & colocalization & $3,782$ & $6,940$ & $30,677$ 
        & $2,238$ 
        & $0.52$ & $0.23$ & $0.40$ \\
        \hline
        & phys. assoc. & dir. iteract. & $2,520$ & $2,610$ & $832$ 
        & $178$ 
        & $0.70$ & $0.93$ & $0.63$ \\
         R. Norvegicus~\cite{de2015structural} & phys. assoc. & colocalization & $2,065$ & $2,757$ & $88$ 
         & $31$ 
         & $0.71$ & $0.91$ & $0.59$ \\
        & dir. iteract. & colocalization & $1,074$ & $991$ & $100$ 
        & $19$ 
        & $0.92$ & $0.91$ & $0.78$ \\
        \hline
    \end{tabular}
    \caption{\textbf{Real-world multiplex networks.} From left to right, we report the name of the dataset and reference where the dataset was introduced, the name of the network layers that compose the multiplex, the number of nodes within the two layers,  the number of edges for each of the two layers, the number of edges shared by the two layers, 
    the modularity of the best Louvain partition of each layer, the normalized mutual information between the two partitions of the two layers.}
    \label{tab:summary_real_multiplexes_pairs}
\end{table*}

\section*{Appendix}

\begin{figure}[!htb]
    \centering
    \includegraphics[width=0.4\textwidth]{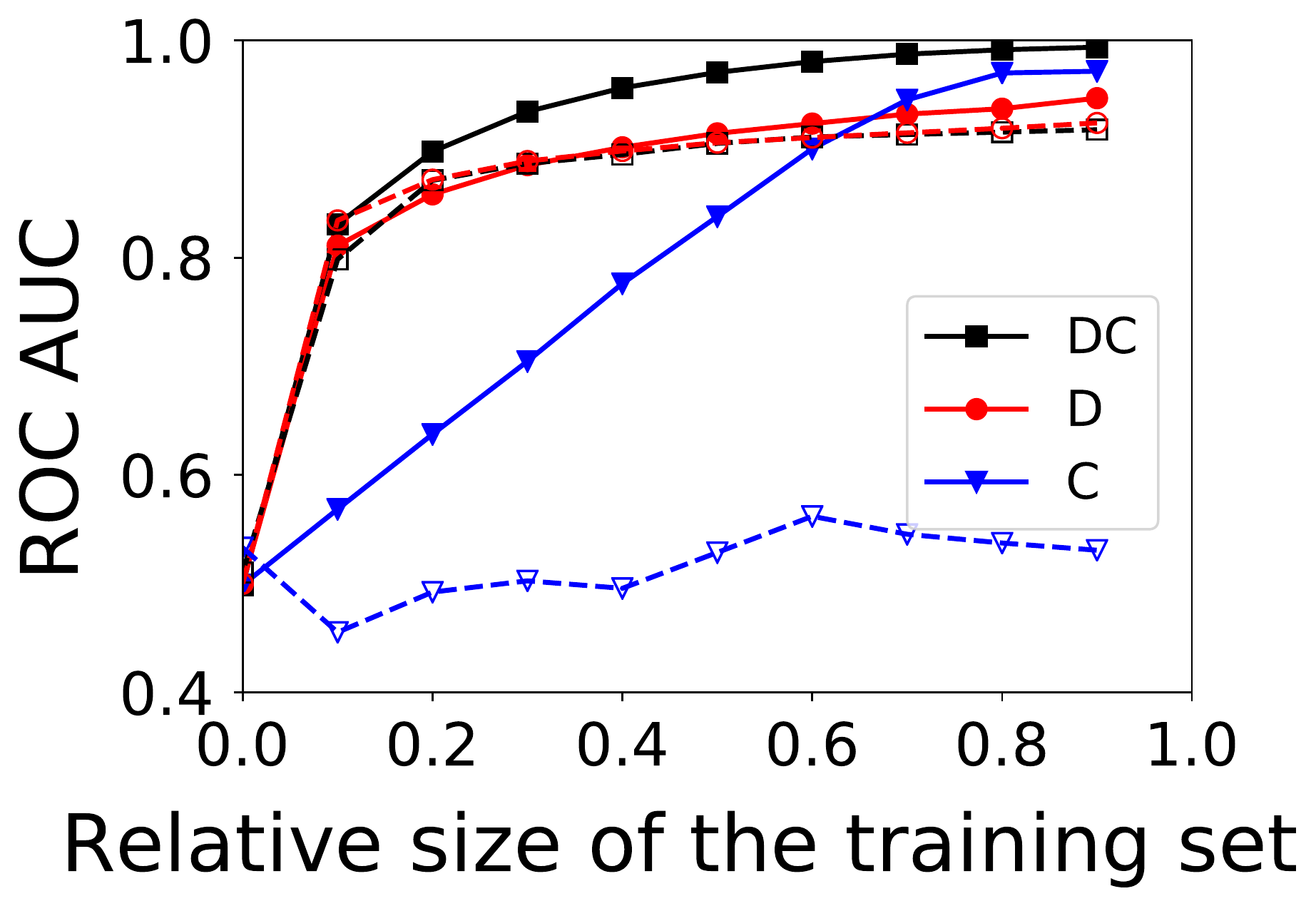}
    \caption{
    \textbf{Reconstruction of synthetic multiplex networks with partial edge information.}
    The figure is the analogue of Fig.~\ref{fig:4}d, but for networks of larger size. We compare the performance of the different classifiers on synthetic graph with built-in community structure (filled symbols / full curves) and without community structure (empty symbols / dashed curves). We plot the performance metric of the degree- and community-based (DC) classifier, the degree-based (D) classifier and the community-based (C) classifier. Graphs with community structure are constructed using the LFR  multiplex model with $N=100,000$, $\gamma=2.1$, $k_{\max} = \sqrt{N}$, $\langle k \rangle = 6$, $\tau=1.0$, and $\mu = 0.1$. We use the configuration model with $N=100,000$, $\gamma = 2.1$, $k_{\min} = 3$, $k_{\max} = \sqrt{N}$ for the graphs with no community structure. The degree sequences and community structures of the layers are uncorrelated.
    }
    \label{fig:7}
\end{figure}

In Fig.~\ref{fig:7}, we repeat the analysis of Fig.~\ref{fig:2}d for networks with size $N=10^5$. For the LFR model, we change the value of the average degree from $\langle k \rangle = 5$ to $\langle k \rangle = 6$ to bypass a convergence issue of the algorithm used to generate network instances. All other parameters are identical to those of Fig.~\ref{fig:2}d.

\begin{figure*}[!htb]
    \centering
    \includegraphics[width=0.95\textwidth]{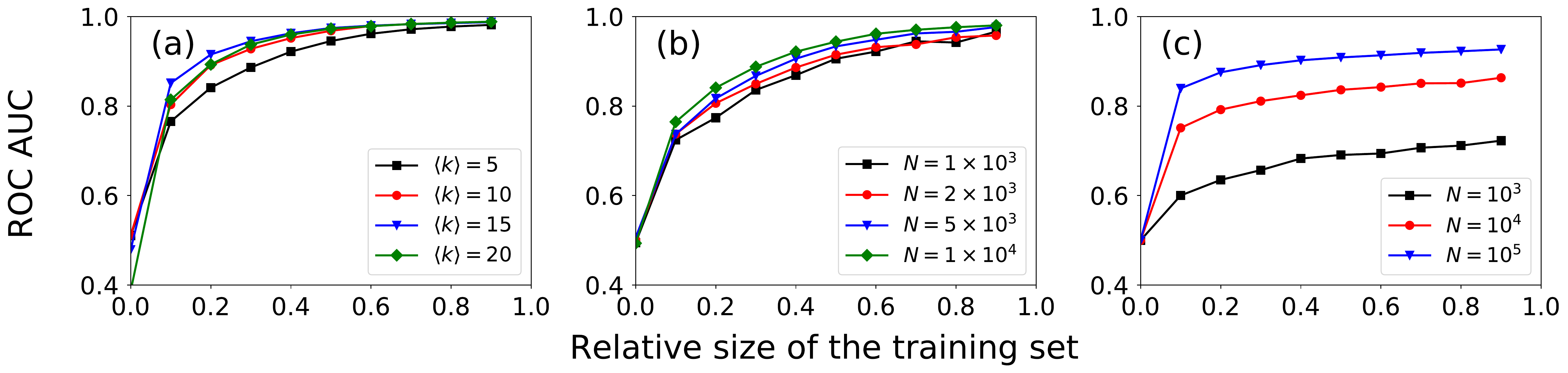} 
    \caption{{\bf Reconstruction of synthetic multiplex networks} (a) We display the ROC AUC of the degree- and community-based (DC) classifier as a function of the the relative size of the training set. We consider  multiplex networks with pre-imposed community structure with size $N=10,000$ and power-law degree distribution $P(k) \sim k^{-\gamma}$, with $\gamma =2.1$. The strength of the community structure is determined by the mixing parameter $\mu = 0.1$. Communities have size obeying a power-law distribution with exponent $\tau = 1.0$. Different curves represent different values of average degree $\langle k \rangle$. Maximum degree is $k_{\max} = \sqrt{N}$. b) Same as in panel (a), but for networks with average degree $\langle k \rangle =5$. Different curves are obtained for different $N$ values. (c) We display the ROC AUC of the degree-based (D) classifier as a function of the the relative size of the training set. Networks  generated according to the
    configuration
    model with $\gamma = 2.1$, $k_{\min} = 3$, and $k_{\max} = \sqrt{N}$. Different curves are obtained for different $N$ values.}
    \label{fig:5}
\end{figure*}

In Fig.~\ref{fig:5}, we show that the performance of the proposed classifier improves
as the size of the network increases, whereas remains almost identical as the average degree of the network is varied. 

\begin{figure*}[!htb]
    \centering
    \includegraphics[width=0.95\textwidth]{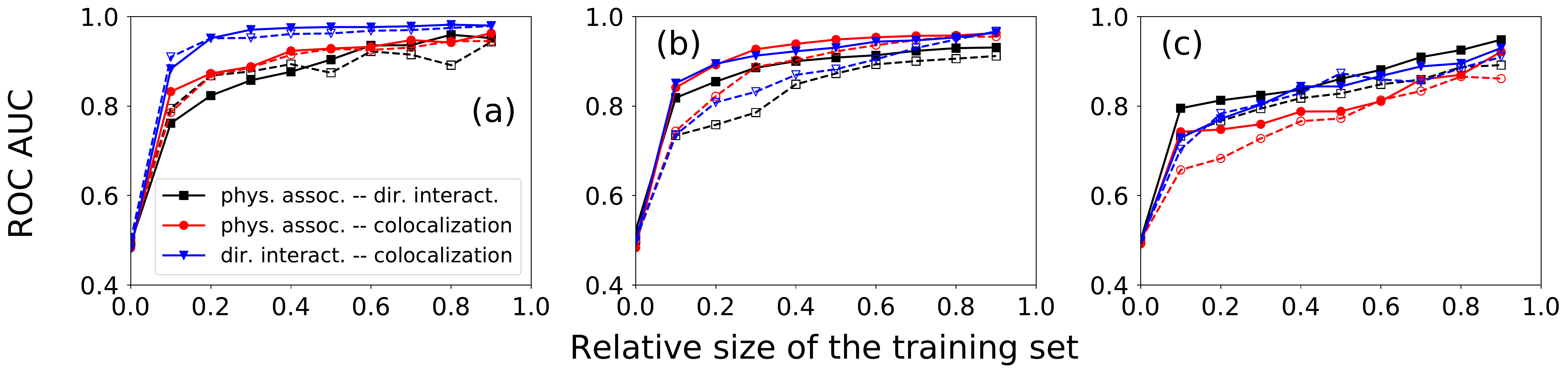} 
    \caption{{\bf Reconstruction of real-world multiplex networks} (a) ROC AUC as a function of the relative size of the training set. We report results for the {\it Saccharomyces Pombe} genetic  multiplex network ~\cite{de2015structural}. Results are averaged over $10$ realizations.
    We compare the results achieved with the degree- and community-based (DC, filled symbols / full curves) classifier with those of the Wu {\it et al.}'s classifier (empty symbols / dashed curves)~\cite{Wu2020}. Results are averaged over $10$ realizations. (b) Same as in panel (a), but for the genetic interactions multiplex network of the {\it Homo Sapiens}~\cite{de2015structural}. Results are obtained over $1$ realization. (c) Same as in panel (a), but for the genetic interactions multiplex network of the {\it Rattus Norvegicus}~\cite{de2015structural}. Results are averaged over $10$ realizations.}
    \label{fig:6}
\end{figure*}

The analysis of three additional real-world multiplex networks is summarized in Fig.~\ref{fig:6}.

\begin{figure*}[!htb]
    \centering
    \includegraphics[width=0.95\textwidth]{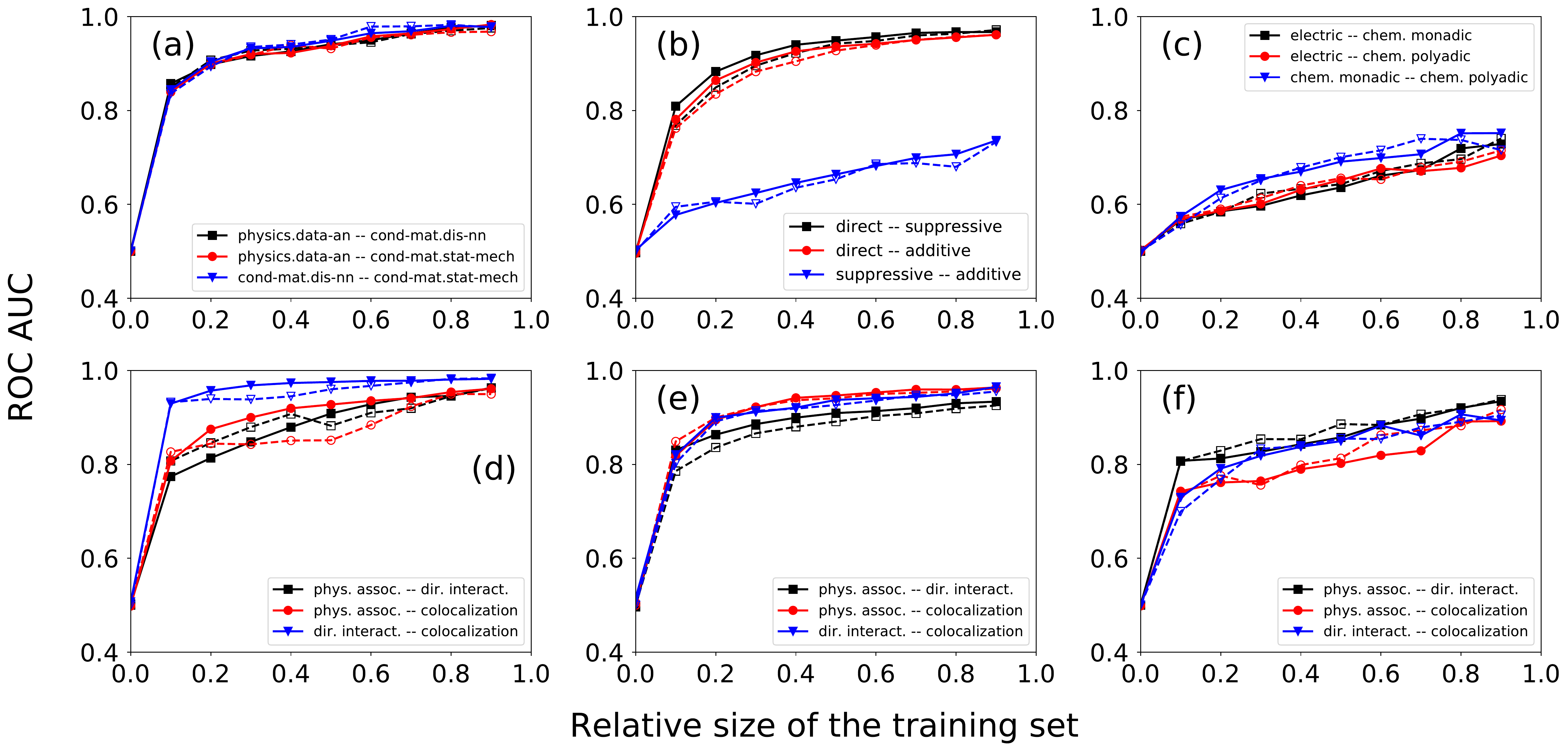} 
    \caption{{\bf Reconstruction of real-world multiplex networks} (a) ROC AUC as a function of the relative size of the training set. We report results for the arXiv  multiplex collaboration network ~\cite{de2015identifying}. 
    We compare the results achieved with the degree- and community-based (DC) relying on two different community detection algorithms: Louvain (filled symbols / full curves)~\cite{blondel2008fast} and Infomap (empty symbols / dashed curves)~\cite{rosvall2008maps}. Results are averaged over $10$ realizations. (b) Same as in panel (a), but for the genetic interactions multiplex network of the {\it Drosophila Melanogaster}~\cite{de2015structural}. Results are averaged over $10$ realizations. (c) Same as in panel (a), but for the {\it Caenorhabditis Elegans} multiplex connectome~\cite{de2015muxviz}. (d) Same as in panel (a), but for the {\it Saccharomyces Pombe} genetic  multiplex network ~\cite{de2015structural}. Results are averaged over $10$ realizations. (e) Same as in panel (a), but for the genetic interactions multiplex network of the {\it Homo Sapiens}~\cite{de2015structural}. Results are obtained over $1$ realization. (f) Same as in panel (a), but for the genetic interactions multiplex network of the {\it Rattus Norvegicus}~\cite{de2015structural}. Results are averaged over $10$ realizations.
    }
    \label{fig:8}
\end{figure*}

In Fig.~\ref{fig:8}, we compare the performance of the DC algorithm based on either Louvain or Infomap. In all networks considered, we do not appreciate an apparent difference in performance due to the specific community detection algorithm leveraged.

\begin{figure*}[!htb]
    \centering
    \includegraphics[width=0.8\textwidth]{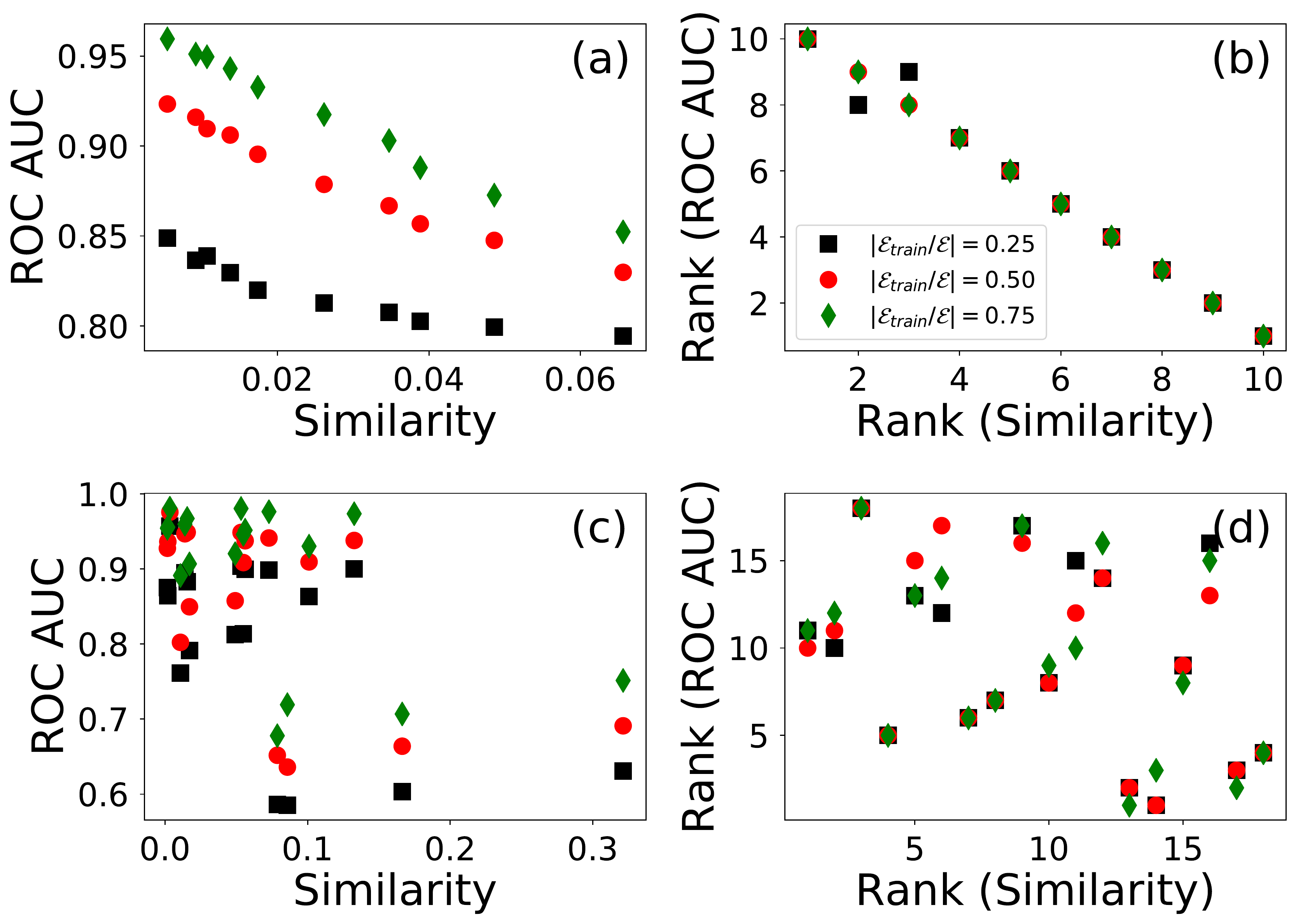}
    \caption{
    {\bf Sensitivity analysis in the reconstruction of multiplex networks with edge overlap.}
        (a) We consider Lancichinetti-Fortunato-Radicchi (LFR) multiplex networks with $N = 10, 000$ nodes, average degree $\langle k \rangle =10$, degree exponent $\gamma=2.1$, maximum degree $k_{\max}=\sqrt{N} = 100$, community distribution exponent $\tau=1$, mixing parameter $\mu=0.1$, and minimum size of the communities $c_{\min}=10$. The two layers are initially identical, then
        nodes of one of the layers are relabeled with probability $q$ to control for the amount of shared edges. For $q=0$, the two layers share all edges; for $q=1$, the fraction of edges shared become minimal. We consider various values of the relabeling probability $q$.
        We report the values of the ROC AUC obtained by using our DC classifier as a function of the Jaccard similarity index of the set of edges of the two layers, i.e., $J = |\mathcal{E}^{(\alpha, \beta)}| / ( |\mathcal{E}^{(\alpha)}|+|\mathcal{E}^{(\beta)}|+|\mathcal{E}^{(\alpha, \beta)}|)$. The different curves correspond to different choices of the relative size of the training set $|\mathcal{E}_{\textrm{train}}| / |\mathcal{E}|$, namely (i) $|\mathcal{E}_{\textrm{train}}| / |\mathcal{E}| = 0.25$, (ii) $|\mathcal{E}_{\textrm{train}}| / |\mathcal{E}| = 0.50$, and (iii) $|\mathcal{E}_{\textrm{train}}| / |\mathcal{E}| = 0.75$, where $\mathcal{E} = \mathcal{E}^{(\alpha)} \cup \mathcal{E}^{(\beta)}$. We perform linear regression of the data points, and obtain the following values of the Pearson's correlation coefficient ($p$-value): (i) $\rho = -0.93 (0.00)$, (ii) $\rho = -0.99 (0.00)$, (iii) $\rho = -0.99 (0.00)$. (b) Same as in panel (a), but we plot the rank position of the data according to the two metrics. We obtain the following values of the Spearman's correlation coefficient ($p$-value):  
        (i) $\sigma = -0.99 (0.00)$, (ii) $\sigma =  -1.00 (0.00)$, (iii) $\sigma = -1.00 (0.00)$. (c) Same as in panel (a), but for the real-world multiplex networks of Fig.~8. We measure: (i) $\rho = -0.54 (0.02)$, (ii) $\rho = -0.55 (0.02)$, (iii) $\rho = -0.56 (0.02)$. (c) Same as in panel (b), but for the real-world multiplex networks of Fig.~8. We measure: (i) $\sigma = -0.34 (0.17)$, (ii) $\sigma =  -0.44 (0.07)$, (iii) $\sigma = -0.42 (0.08)$.
    }
    \label{fig:9}
\end{figure*}

Finally in Fig.~\ref{fig:9},
we study the correlation between the performance of the DC classifier in the reconstruction of a given multiplex and the 
fraction of overlapping edges among the layers of the multiplex.
We consider synthetic (Figs.~\ref{fig:9}a and ~\ref{fig:9}b) and real-world (Figs.~\ref{fig:9}c and ~\ref{fig:9}d) networks. In the case of real networks,
we exclude on purpose the London transportation networks from the analysis
because of their peculiarity of being graphs embedded in physical space. 
As the plot shows, there is a clear relationship between the value of the ROC AUC and the relative fraction of overlapping edges among the layers of the multiplex for synthetic systems. The value of the ROC AUC is inversely proportional to the fraction of edges shared between layers, and disregarded by the classifier. This finding is also confirmed in real networks, although the dependence among edge overlap and reconstruction performance appears less apparent than in synthetic multiplexes.

%

\end{document}